\documentclass{jfm}

\usepackage{graphicx}
\usepackage{epic,eepic}
\usepackage{graphics}
\usepackage{epsfig}
\usepackage{subfigure}
\usepackage{natbib}
\usepackage{amssymb,amsmath,color,epic,eepic,graphicx}
\usepackage{hyperref}
\usepackage{float}

\usepackage{bm}
\usepackage{bbm}
\usepackage{times}

\newcommand{\be}{\begin{equation}}
\newcommand{\ee}{\end{equation}}
\newcommand{\bea}{\begin{eqnarray}}
\newcommand{\eea}{\end{eqnarray}}

\newcommand{\uv}{\bm{u}}

\newcommand{\Rtau}{Re_{\tau}}

\newcommand{\ut}{u_{\tau}}

\DeclareGraphicsExtensions{.jpg, .eps}

\title{Law of the wall in an unstably stratified turbulent channel flow}

\author[A. Scagliarini, H. Einarsson, A. Gylfason, F. Toschi]
{A.\ns S\ls C\ls A\ls G\ls L\ls I\ls A\ls R\ls I\ls N\ls I$^{1,2,3}$ 
H.\ns E\ls I\ls N\ls A\ls R\ls S\ls S\ls O\ls N$^{2}$ \break%
A.\ns G\ls Y\ls L\ls F\ls A\ls S\ls O\ls N$^{2}$ 
F.\ns T\ls O\ls S\ls C\ls H\ls I$^{3,4,5}$\break 
}
\affiliation{$^{1}$ Department of Physics and INFN, University of Rome
  ``Tor Vergata'', Via della Ricerca Scientifica 1, 
00133 Rome, Italy\\ 
$^{2}$ School of Science and Engineering, Reykjavik University, Menntavegur 1, IS-101 Reykjavik, Iceland\\
$^{3}$ Department of Mathematics and Computer Science, Eindhoven
University of Technology, The Netherlands\\
$^{4}$ Department of Applied Physics, Eindhoven
University of Technology, The Netherlands\\
$^{5}$ IAC-CNR, Via dei Taurini 19, 00185 Rome, Italy
}

\pubyear{2014}
\volume{nr}
\pagerange{a--b}

\date{Nov 1, 2014}

\begin{document}
\maketitle
\begin{abstract}
We perform direct numerical simulations of an unstably stratified turbulent channel flow to address the effects of 
buoyancy on the boundary layer dynamics and mean field quantities. We systematically span a range of parameters 
in the space of friction Reynolds number ($\Rtau$) and Rayleigh number ($Ra$). 
Our focus is on deviations from the logarithmic
 law of the wall due to buoyant motion. The effects of convection in the relevant 
ranges are discussed providing measurements of mean profiles of velocity, temperature and Reynolds stresses as well as
 of the friction coefficient. A phenomenological model is proposed and shown to capture the observed deviations of 
the velocity profile in the log-law region from the non-convective case. 
\end{abstract}
\section{Introduction}
Wall bounded turbulence is highly relevant to a variety of engineering systems and naturally occurring flows. As such, 
the importance of the fundamental understanding of the generation of turbulent fluctuations by fluid-solid 
interactions in the boundary layers and their transport and interaction with the bulk mean flow \citep{Pope} 
is evident. In many geophysical and industrial flows, shear flows and wall turbulence are subjected to either 
stable or unstable thermal stratifications. When
  gravity and/or temperature differences are strong enough, buoyancy
  forces may alter significantly the boundary layer dynamics. 
Stable stratification is known to inhibit the bursting phenomenon, i.e. depleting the transport of 
turbulent fluctuations from the wall to the bulk, thus reducing the turbulent drag and increasing the mean flow 
velocity. Many studies have been devoted to this situation when the stratified scalar field is passive 
\citep{JohanssonWikstrom,PapaHanratty} and active \citep{ArmenioSarkar,GerzEtAl,IidaEtAl,VillalbaAlamo} and even in 
presence of non-Oberbeck-Boussinesq effects \citep{ZontaEtAl}.

The unstable configuration is also relevant in a variety of instances, such as the physics of the 
atmospheric layer superposed to an over-heated ground (as, e.g., in summer days, determining
the generation of thermo-convective storms \citep{Bluestein}). Early theoretical results 
on turbulent boundary layers under unstable thermal stratification date back to \citet{Prandtl32} who pioneered a 
mixing-length based approach which inspired the later work of \citet{Obukhov46} anticipating the Monin-Obukhov 
similarity theory (henceforth MO54) \citep{MO54}. These pioneering works
  have boosted a number of experimental 
  \citep{Lenschow,BusingerEtAl,KaimalEtAl,HuntEtAl} and numerical
  \citep{Deardroff72,Deardroff74,Moeng} works, as well as further theoretical developments based
  either on $k-\varepsilon$ theories \citep{MellorYamada1,MellorYamada2} or on the Rapid
Distortion Theory (RDT) \citep{Dubrulle1,Dubrulle2} on the dynamics of
planetary boundary layers (PBL). Also, in another seminal paper, \citet{KaderYaglom} revised MO54 and
compared it with experimental data.
\citet{LumleyEtAl} proposed an eddy-damped quasi-Gaussian closure able to predict the inversion regions of heat flux 
profiles in strongly buoyant sheared boundary layers. Numerical results \citep{IidaKasagi} and experimental flume 
measurements (combined with a spectral equation model) \citep{KomoriEtAl} showed that natural thermal convection 
affects the mechanisms of momentum and heat transport from the wall and tends to flatten the velocity profile in the 
bulk; these observation were confirmed by large-eddy simulations of a variable density fluid \citep{ZainaliLessani}, 
although with non-Boussinesq effects, as, e.g. profile asymmetries 
emerging at large stratifications.

In this work we perform direct numerical simulations based on the lattice Boltzmann (hereafter LB) method of an 
unstably (thermally) stratified turbulent channel flow. Our focus is on identifying the effects of buoyancy on the 
channel flow structure by comparing profiles of mean fields and fluctuations over a wide range of parameters with a 
pure (unstratified) channel flow. Our results show a decreased fluid throughput due to a strongly flattened velocity 
profile, which could, however, be fitted by a log-law with coefficients depending on the input controlling parameters 
(friction Reynolds number $Re_{\tau}$ and Rayleigh number $Ra$, defined below), as also derived theoretically.
\section{Numerical method and simulation details}
We simulated a fluid enclosed between two walls, kept at fixed temperatures $T_H$ at $y=0$ and 
$T_C = T_H - \Delta$ at $y=2H$, and driven by a constant body force
mimicking an imposed pressure gradient along 
the streamwise direction $\hat{x}$.
The equations of motion are the incompressible Navier-Stokes equation
for the fluid velocity field $\uv(\bm{x},t)$ in the Boussinesq
approximation (namely, the density is assumed to be constant
$\rho=\rho_0$, but for
the linearised buoyancy term)
\begin{equation} \label{eq:NS} 
\partial_t \uv + \uv \cdot \bm{\nabla}\uv= - \bm{\nabla} P +
\nu \triangle \uv - \beta\bm{g}(T-T_m) + \bm{F},
\end{equation} 
($T_m = (T_H+T_C)/2$ being the mean temperature) coupled with the advection-diffusion equation for 
the temperature field 
$T(\bm{x},t)$
\begin{equation} \label{eq:temp} 
\partial_t T + \uv \cdot \bm{\nabla} T = \alpha \triangle T.
\end{equation}
In the above equations, $P$ is the pressure field (rescaled by
$\rho_0$), $\bm{g} = -g \hat{y}$ is the acceleration of gravity and
$\bm{F} = (\ut^2/H)\hat{x}$ is the acceleration due to the constant
body force ($\ut$ being the friction velocity); $\nu$, $\alpha$ and 
$\beta$ are the kinematic viscosity, the thermal diffusivity and thermal expansion coefficient, respectively.
As a numerical scheme, we adopted a 3d LB algorithm \citep{Benzi92,Chen98,Aidun10} with 
two probability densities (for density/momentum and for temperature, respectively) \citep{He98}. 
The method has been extensively used to study both thermal
convection \citep{BenziToschiTripiccione,Calzavarini}  and turbulent channel
flow \citep{ToschiEtAl99,ToschiEtAl2000,BiferaleEtAl2002}; in
particular \citet{LavezzoEtAl} validated the code
and tested grid resolutions against previous studies with
different numerical methods.

We used a computational grid $L \times 2H \times W$ of $256 \times 128 \times 128$ lattice points, with each run 
longer than $3 \times 10^6$ time steps (in LB units), in such a way to achieve 
statistically steady states of $\sim 200 T_L$ ($T_L$ being the large-scale eddy turnover time). 
From equations  (\ref{eq:NS}) and (\ref{eq:temp}) two dimensionless groups can be identified giving rise to two 
parameters which control the dynamics, namely the shear (or friction) Reynolds number,
$$
\Rtau = \frac{\ut H}{\nu} 
$$
and the Rayleigh number 
$$
Ra = \frac{\beta g \Delta (2H)^3}{\nu \alpha}
$$
quantifying, respectively, the strength of the pressure induced shear and of the buoyancy with respect to viscous 
dissipation.

We performed turbulent channel flow simulations with $\Rtau \in [46,
205]$; for each $\Rtau$ we tuned the gravity (at fixed temperature
jump), whence the buoyancy, spanning the range $Ra \in [0, 1.3 \times
10^7]$. The Prandtl number $Pr = \nu/\alpha$ is equal
  to one in all cases presented.
\section{Results} 
\begin{figure}
\begin{center}
\includegraphics[scale=0.5]{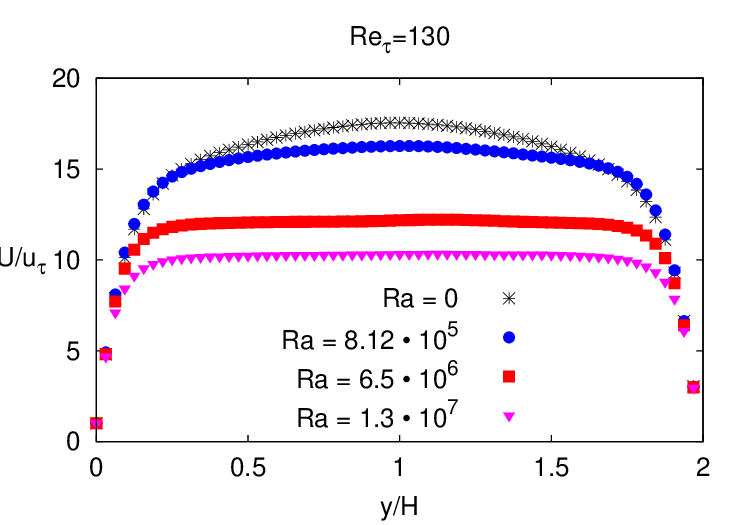}
\includegraphics[scale=0.5]{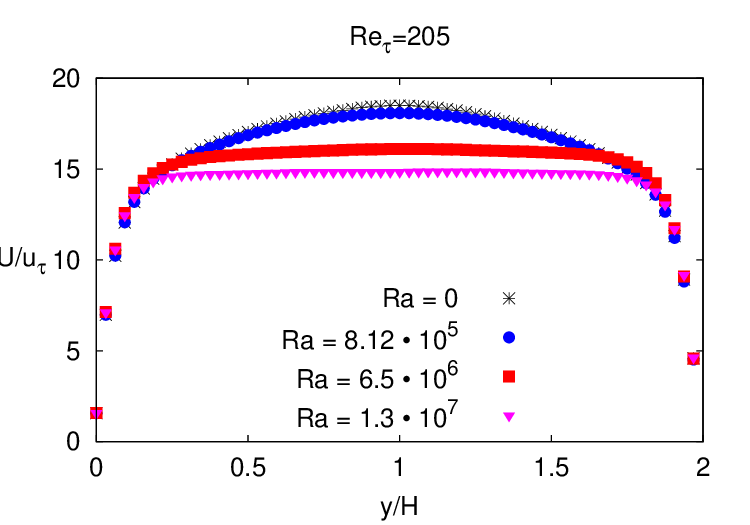}
\caption{Mean profiles of the streamwise velocity $U(y)$, normalized with the friction velocity $u_{\tau}$,
for various Rayleigh numbers 
($Ra=0$, $Ra=8.12\times 10^5$, $Ra=6.5\times 10^6$ and $Ra=1.3\times 10^7$) at fixed shear Reynolds number $\Rtau$. 
Left: $\Rtau=130$, Right: $\Rtau=205$. The data have been plotted every two points
   (i.e. lattice spacings) for the sake of clarity of visualisation}
\label{02U_mean_norm}
\end{center}
\end{figure}

\begin{figure}
\begin{center}
\includegraphics[scale=0.5]{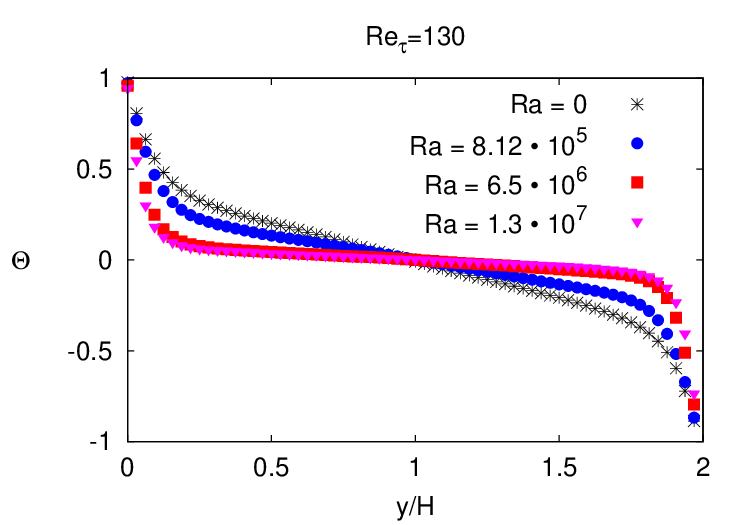}
\includegraphics[scale=0.5]{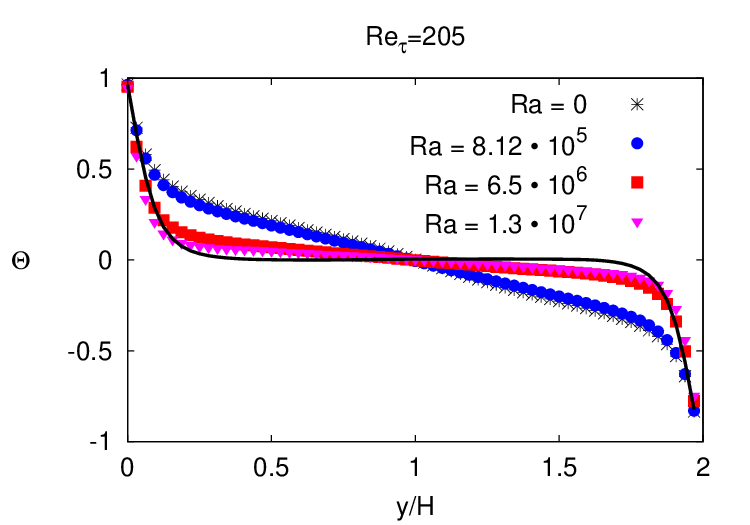}
\caption{Mean profiles of the normalized temperature $\Theta = 2(T - T_m)/\Delta$ 
as a function of the height $y$ for various Rayleigh numbers 
($Ra=0$, $Ra=8.12\times 10^5$, $Ra=6.5\times 10^6$ and $Ra=1.3\times 10^7$) at fixed shear Reynolds number $\Rtau$. 
Left: $\Rtau=130$, Right: $\Rtau=205$; the solid line represents the profile for the purely
  convective Rayleigh-B\'enard case (i.e. $Re_{\tau} = 0$) with
  $Ra=8.12 \times 10^5$. The data have been plotted every two points
   (i.e. lattice spacings) for the sake of clarity of visualisation}
\label{03T_mean_norm}
\end{center}
\end{figure}

\begin{figure}
\begin{center}
\begin{minipage}[b]{0.45\linewidth}
\centering
\includegraphics[width=\textwidth]{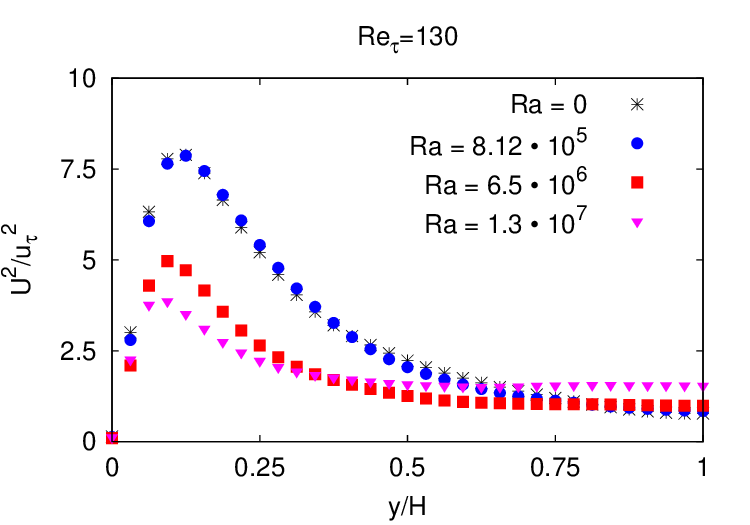}
\end{minipage}
\begin{minipage}[b]{0.45\linewidth}
\centering
\includegraphics[width=\textwidth]{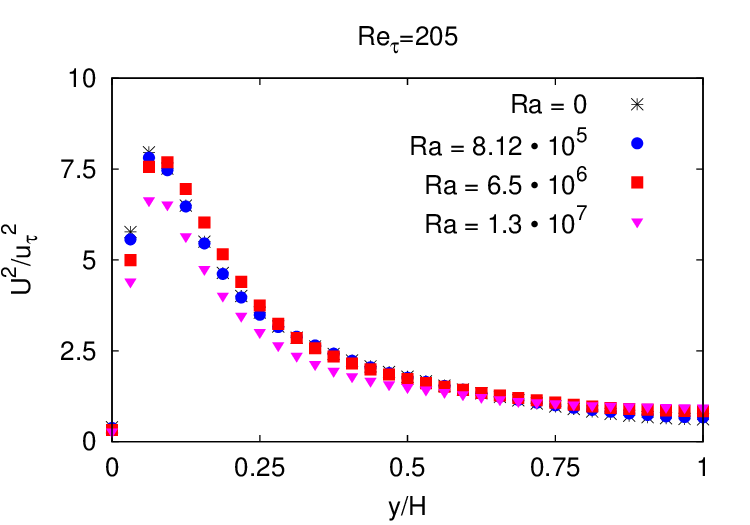}
\end{minipage}
\begin{minipage}[b]{0.45\linewidth}
\centering
\includegraphics[width=\textwidth]{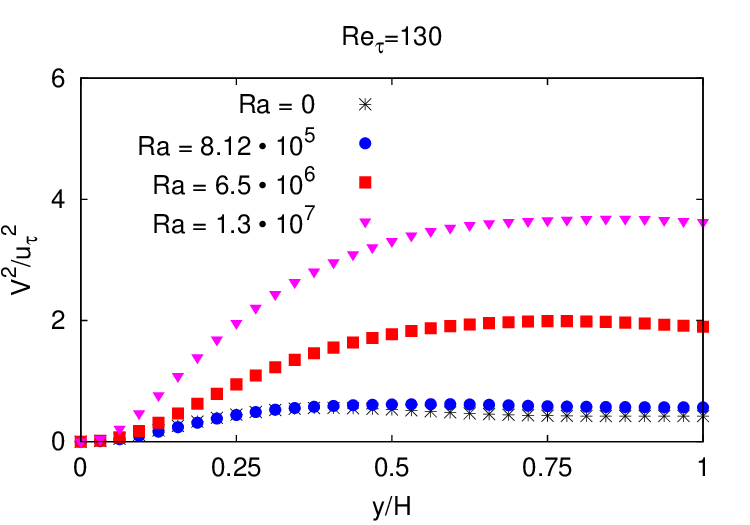}
\end{minipage}
\begin{minipage}[b]{0.45\linewidth}
\centering
\includegraphics[width=\textwidth]{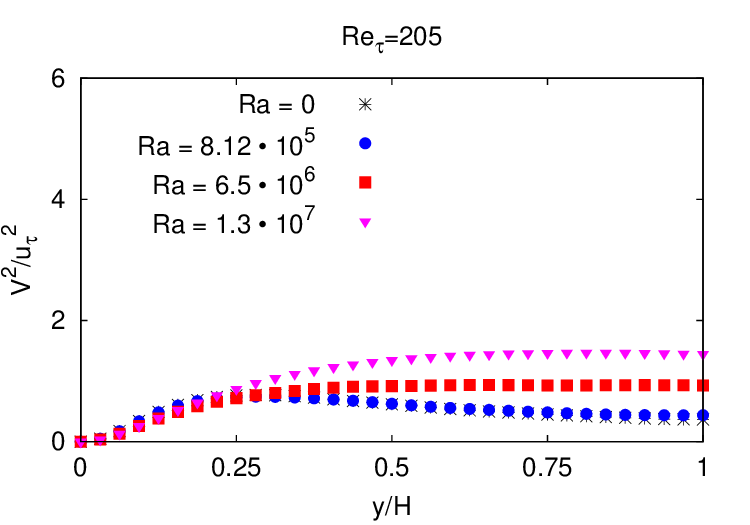}
\end{minipage}
\caption{Profiles of longitudinal $U^2 =
 \overline{\tilde{u}_x^2}$ and transverse $V^2 =
 \overline{\tilde{u}_y^2}$ mean squared components of the fluctuating
 velocity field, normalized by the
 friction velocity $u_{\tau}$ for various Rayleigh numbers ($Ra=0$, $Ra=8.12\times 10^5$, $Ra=6.5\times 10^6$ and $Ra=1.3\times 10^7$) at 
fixed shear Reynolds number $\Rtau$.}
\label{04Re_stress}
\end{center}
\end{figure}

\begin{figure}
\begin{center}
\includegraphics[width=0.6\textwidth]{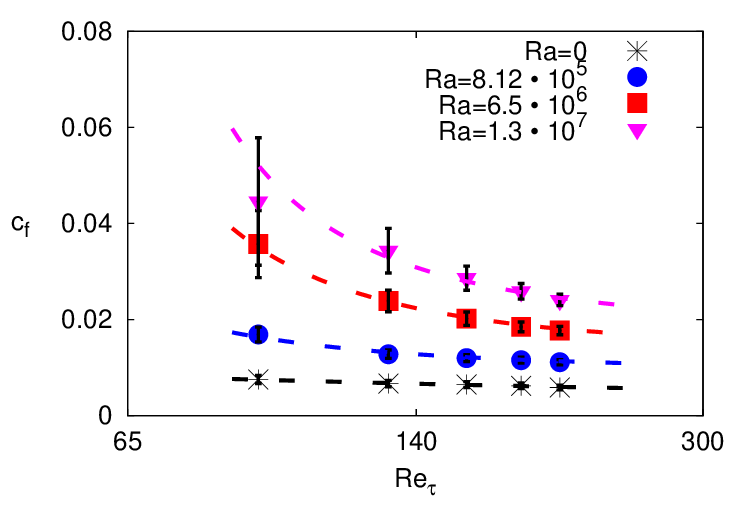}
\caption{The skin-friction coefficient $c_f=\tau_w/(1/2 U_0^2)$ as
  a function of the shear Reynolds number $Re_{\tau}$ 
 for various Rayleigh numbers $Ra$. The dashed lines are predictions obtained by approximating the mean velocity by a 
log-law over the whole channel; for the stratified cases we made use
of equations (\ref{eq:tLog-Law})-(\ref{eq:Kc}) with the same fitting
parameter $C_s=2.5$ for all $Ra$ (see section \ref{Modelling}.)} 
\label{06c_f}
\end{center}
\end{figure}

Mean profiles of the streamwise velocity $U(y)= \overline{u_x}$ (the overline indicates here and henceforth averaging 
over planes parallel to the walls $\overline{(\cdot)} = (L_xL_z)^{-1}\int (\cdot) dxdz$)
in the channel are shown in figure \ref{02U_mean_norm}. 
For each Reynolds number ($Re_\tau=130$ and $Re_\tau=205$) we show data from simulations at  
three different Rayleigh numbers ($Ra=8.12\times 10^5$, $Ra=6.5\times 10^6$ and $Ra=1.3\times 10^7$),
besides the unstratified channel flow ($Ra=0$).
An evident effect of thermal stratification is a decrease of the centreline velocity and a flattening of the 
profiles at increasing $Ra$; indeed, mixing between the bulk and the boundary layer regions due to wall-normal 
thermal fluctuations (or ``plumes'') results in an increase of the effective wall drag \citep{HattoriEtAl}.
Such effects are more pronounced for lower $\Rtau$, as it clearly appears by comparison of left and right panels of 
figure \ref{02U_mean_norm}. These observations will be discussed  more quantitatively under the light 
of the modelling in the next section.
Figure \ref{03T_mean_norm} shows the mean temperature profile with respect to $y$, for the same cases 
displayed in figure \ref{02U_mean_norm}. 
Here, we note the bending of the thermal profiles in the bulk with
increased shear Reynolds number, 
in contrast to the thermal shortcut observed in pure
turbulent Rayleigh-B\'enard (solid line in the right
  panel)  \citep{RB1,RB2}, owing to the destruction or sweeping of the
coherent plumes rising from the wall surfaces \citep{ScagliariniEtAl}.   
Longitudinal and transverse mean squared components of the fluctuating velocity field
(hereafter $\tilde{u}_i = u_i - \overline{u}_i$) are shown in figure \ref{04Re_stress} for the same cases as in figure 
\ref{02U_mean_norm}  (normalized with the friction velocity). When the fluctuating quantities are concerned, 
most notably the lateral component is increased in the bulk region, and as the Rayleigh number is increased, the 
magnitude becomes comparable or greater than the stream-wise component. The near wall regions are also affected, 
primarily in the streamwise component, where an increase and a subsequent decrease in magnitude is observed as the 
Rayleigh number is increased. 
In order to quantify the overall effect of the thermal forcing on the channel flow, figure \ref{06c_f} shows the 
skin-friction coefficient $c_f=\tau_w/(1/2 U_0^2)$ as a function of shear Reynolds number $\Rtau$. The increased 
drag due to the thermal field is evident, resulting in higher than usual friction coefficient. The effect is reduced 
asymptotically, as the Reynolds number is increased for a given Rayleigh number, emphasizing that shear becomes the 
dominant source of turbulence at sufficiently high Reynolds numbers.
\section{Modification of the law of the wall by buoyancy} \label{Modelling}
\begin{figure}
\begin{center}
\includegraphics[scale=0.8]{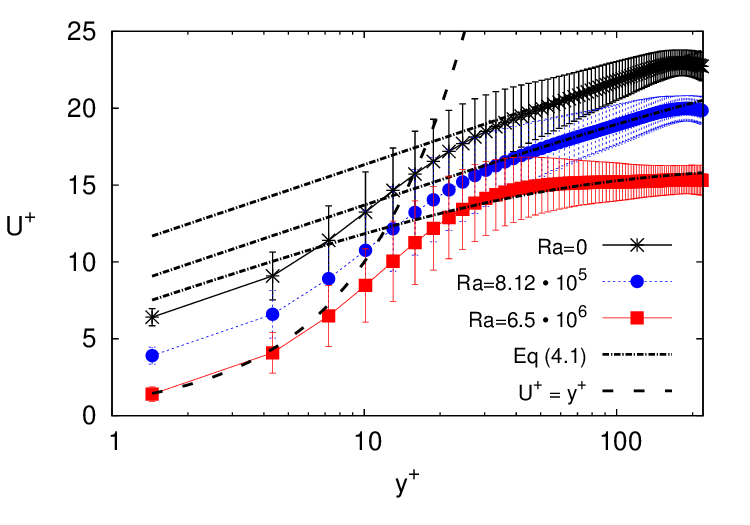}
\caption{Lin-log plot of the velocity profiles in wall units, for
  shear Reynolds number $Re_{\tau}=185$ and  Rayleigh
  numbers $Ra=0$ ($\ast$), $Ra=8.12 \times 10^5$
  (\textcolor{blue}{$\bullet$}) and $Ra=6.5 \times 10^6$
  (\textcolor{red}{$\blacksquare$}); the data are shifted vertically
  for clarity of visualisation.
 The dashed lines indicate the linear law, valid in the viscous
 layer, $U^+ = y^+$, while the dash-dotted ones correspond to the
 prediction (\ref{eq:tLog-Law})-(\ref{eq:Kc}) with $\kappa = 0.42$,
 $C_S=2.5$ and $B \in [5.8, 6.7]$.}
\label{Fig:vel-log-law}
\end{center}
\end{figure}

Inspection of the profiles of the mean quantities and the fluctuations showed that buoyancy alters the channel flow structure. We
 now focus on the mean velocity profiles. In figure \ref{Fig:vel-log-law} we report the lin-log plot of profiles, in wall
 units, for $\Rtau=185$ and for various $Ra$. Convective motion, realised in the form of thermal plumes rising from the
 walls to the bulk, disturbs the coherence of the channel flow, resulting in a source of drag which decreases the mean 
velocity in the channel and hence the mass throughput, as well as flattening the velocity profile.  
In order to quantify these observations, in the following we attempt
to generalize von K\'arm\'an's law of the wall accounting for
buoyancy effects. To this aim, we propose a simple model constructed
from conservation laws and
phenomenological arguments, which predicts that the viscous buffer is
insensitive to buoyancy (i.e. $U^+ = y^+$), while in the log-law 
region the following relation
\begin{equation} \label{eq:tLog-Law}
U^+(y^+) = \kappa^{-1}\log\left(\frac{y^+}{1+\kappa_C y^+} \right) + B,
\end{equation}
holds (quantities have been expressed in wall units $U^+ \equiv U/\ut$ and $y^+ \equiv y/\delta$, 
with $\delta = \nu/\ut$). In equation (\ref{eq:tLog-Law}), $\kappa$ is
the von K\'arm\'an constant, $B$ is an integration constant and $\kappa_C$ is given by
\begin{equation} \label{eq:Kc}
\kappa_{C}(Ra,\Rtau) = C_S \frac{Ra}{\Rtau^4 Pr^2};
\end{equation}
$C_S$ is a phenomenological parameter. All profiles in figure
\ref{Fig:vel-log-law} agree reasonably, within error bars, with our predictions: 
the data collapse on the function $U^+=y^+$ for $y^+ < 10$ (dashed
lines) while in the logarithmic layer they can be fitted with 
equation (\ref{eq:tLog-Law}) (dash-dotted lines), whose derivation is detailed in what
follows.\\
Applying the Reynolds decomposition $\bm{u}(\bm{x},t)=U(y) \hat{x} + \tilde{\bm{u}}(\bm{x},t)$ 
for the velocity field and averaging the Navier-Stokes equation for
the $\hat{x}$ velocity component, we get, upon integration in $y$, an 
exact relation \citep{Pope} between the mean shear $S(y) = dU(y)/dy$
and the Reynolds stress 
$\tau_{xy} (y) = -\overline{\tilde{u}_x \tilde{u}_y}$ namely:
\begin{equation} \label{eq:aveNS1}
\nu S(y) + \tau_{xy} (y) = \ut^2 \left(1 - \frac{y}{H}
    \right),
\end{equation} 
or, in wall units:
\begin{equation} \label{eq:aveNS2}
 S^+(y^+) + \tau^+_{xy} (y^+) = \left( 1 - \frac{y^+}{Re_{\tau}} \right).
\end{equation} 
Notice that equations (\ref{eq:aveNS1})-(\ref{eq:aveNS2}) remain valid also in the presence of buoyancy, since the forcing term 
$\beta g (T-T_m) \hat{y}$ does not act along the streamwise direction.
For very large $Re_{\tau}$ (in principle $Re_{\tau} \rightarrow
\infty$), equation (\ref{eq:aveNS2}) gives $S^+ + \tau^+_{xy} = 1$;
close to the wall (for $y^+ < 10$), where viscous
terms dominate, $S^+ \gg \tau^+_{xy}$ whence $S^+ \approx 1$ which
implies $U^+ \simeq y^+$, as anticipated above. Conversely, away from
the wall (typically for $y^+ \stackrel{>}{\sim} 30$), turbulent
fluctuations dominate over viscous processes and $\tau^+_{xy} \gg S^+$,
i.e. \citep{LvovEtAl}
\begin{equation} \label{eq:tauconst}
\tau^+_{xy} \approx const;
\end{equation} 
such result alone, however, 
does not give any further insight on the behaviour of the velocity
profile in the corresponding region of the channel.
To this aim, we need to consider the
budget equation of turbulent kinetic energy $E_K(y) =
\overline{|\tilde{\bm{u}}|^2}/2$ \citep{Pope}. 
In the log-law layer the energy production $\mathcal{P} =
\tau_{xy} (y)S(y) + \beta g Q(y)$ (where $Q(y)=\overline{u_y \theta}$
is the turbulent heat flux and $\theta$ are the temperature
fluctuations) is balanced by
dissipation $\varepsilon(y) = 2\nu \overline{(\partial_i
  \tilde{u}_j)^2}$ \citep{Pope}; the latter cannot be calculated exactly but
only inferred by means of phenomenological arguments: for large
Reynolds number and outside the viscous boundary layer, the
dissipation equals the turbulent energy flux, which can be estimated
as $E_K(y)/\tau(y)$, where $\tau(y) \propto y/\sqrt{E_K(y)}$ is the typical 
eddy turn-over time at $y$ \citep{LvovEtAl}. Thus, the energy balance
equation reads:
\begin{equation} \label{eq:clos1}
\frac{a E_K^{3/2}(y)}{y} = \tau_{xy} (y)S(y) + \beta g Q(y),
\end{equation}
where $a$ is a non-dimensional parameter of order unity. 

\begin{figure}
\begin{center}
\includegraphics[width=0.6\textwidth]{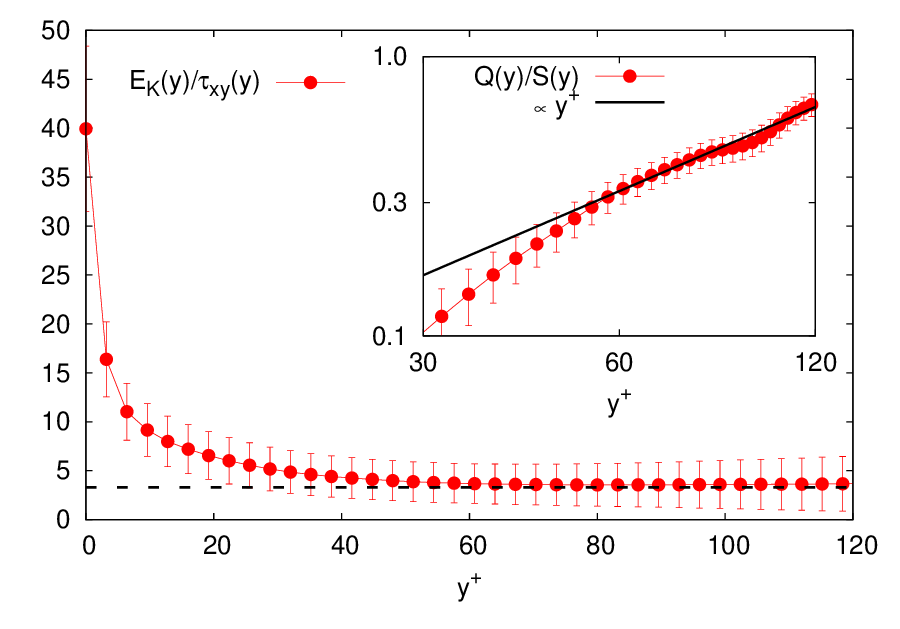}
\caption{Main panel: Ratio of the turbulent kinetic energy $E_K$ over
  the Reynolds stress $\tau_{xy}$ (for the case $\Rtau=205$,
  $Ra = 8.12 \times 10^5$): the plateau attained for $y^+ \stackrel{>}{\sim} 40$
  suggests that the two quantities are
  proportional to each other in the log-law region. Inset: Ratio of
  the turbulent heat flux $Q(y)$ over the mean
  shear $S(y)$; the solid line
  indicates a linear dependence on $y^+$ and represents a numerical
  check of the validity of the assumptions leading to equation (\ref{eq:ml3}).}
\label{Fig:checkDNS}
\end{center}
\end{figure}

\begin{figure}
\begin{center}
\includegraphics[width=0.6\textwidth]{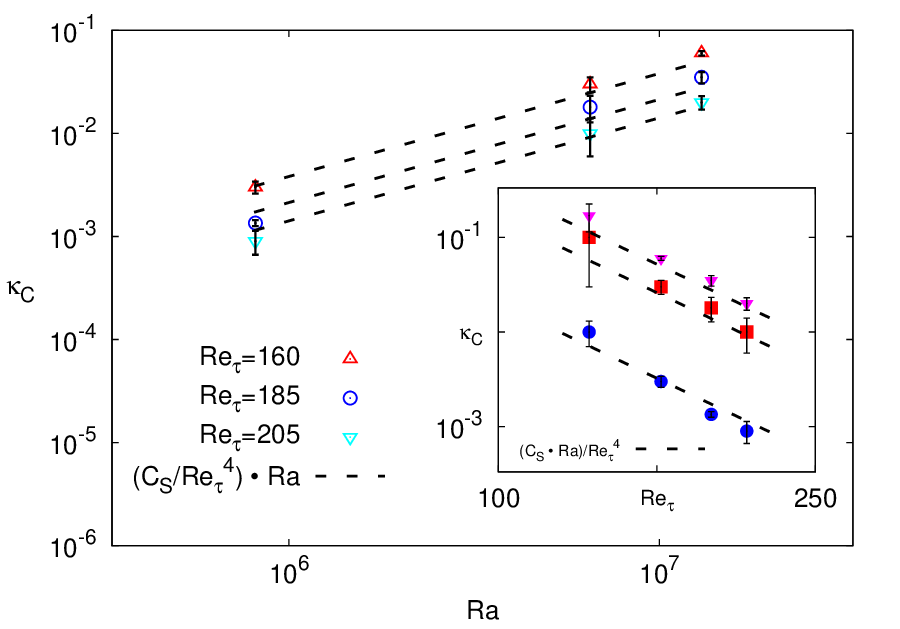}
\caption{Plots of the parameter $\kappa_C(Ra,\Rtau)$ of the model as function of $Ra$ for fixed $\Rtau$'s (main panel) 
and as function of $\Rtau$ for fixed $Ra$'s (inset),
whose values are $Ra=8.12 \cdot 10^5$
  (\textcolor{blue}{$\bullet$}), $Ra=6.5 \cdot 10^6$
  (\textcolor{red}{$\blacksquare$}) and $Ra = 1.3 \cdot 10^7$
  (\textcolor{magenta}{$\blacktriangledown$}). 
The dashed lines represent equation (\ref{eq:Kc}) with constant $C_S$ 
fixed at $C_S=2.5$ for all cases.}
\label{Fig:tLogLawConstant}
\end{center}
\end{figure}

Equation (\ref{eq:clos1}) is not yet closed: a relation between $E_K(y)$ and $\tau_{xy} (y)$ is required. Dimensional analysis 
suggests that the two quantities should be proportional to each other,
i.e. $E_K(y) \propto \tau_{xy}(y)$ \citep{LvovEtAl}; such assumption
can be readily tested in the numerical simulations: in the main panel
of figure \ref{Fig:checkDNS} we show that the ratio
$E_K(y)/\tau_{xy}(y)$, indeed, approaches a constant value for $y^+ \stackrel{>}{\sim} 40$.
We can then rewrite (\ref{eq:clos1}) as (in view also of
(\ref{eq:tauconst})) 
\begin{equation} \label{eq:clos2}
\frac{a (E_K)^{3/2}}{y} = c E_K S(y) + \beta g Q(y),
\end{equation}
with $c$ yet another dimensionless number.
The energy production by buoyancy term, viz. the heat flux, needs also to be modelled; although more refined closures 
can be employed \citep{JohanssonWikstrom,HattoriEtAl} involving tensorial eddy diffusivities and coupling with 
gradients of the temperature field in all directions, we adopt a
simple standard mixing length {\it ansatz}, i.e.
\begin{equation} \label{eq:ml}
Q(y) = \overline{u_y \theta} =- \ell_m^2 \partial_y
\overline{U} \partial_y \overline{T}.
\end{equation}
A first order closure like (\ref{eq:ml}) must not be expected to work well for second order quantities like temperature
 fluctuations, Reynolds stresses, etc, but, as we will show, mean streamwise velocity profiles are satisfactorily 
reproduced through such model; according to Prandtl's hypothesis the mixing length $\ell_m$ is proportional to the 
distance from the wall, i.e. $\ell_m \propto y$, whence
\begin{equation} \label{eq:ml2}
Q(y) = \overline{u_y \theta} = -b y^2 \partial_y
\overline{U} \partial_y \overline{T},
\end{equation}
where $b$ is a numerical constant. 
The mean temperature gradient will, in principle, depend itself on the
shear profile; however, in a perturbative spirit, we postulate here,
for simplicity, a logarithmic form, such that 
\begin{equation} \label{eq:meanTprof} 
\frac{d \overline{T}}{dy} = - \frac{T_*}{y},
\end{equation}
where $T_*=\frac{\alpha \Delta}{2H \, u_{\tau}}$ is a scale for the
dynamic temperature. The functional form (\ref{eq:meanTprof}) somehow
interpolates between the passive scalar case \citep{JohanssonWikstrom}
and the natural convection \citep{AhlersEtAl} and
  it is in agreement with a theoretical prediction based on RDT \citep{Dubrulle2}. Inserted in the
expression for the heat flux (\ref{eq:ml2}), equation (\ref{eq:meanTprof}) gives
\begin{equation} \label{eq:ml3}
Q(y) \propto b y^2 S(y) \frac{T_*}{y} = b^{\prime} y S(y) T_*.
\end{equation}
For the sake of validation of our arguments, we check equation
(\ref{eq:ml3}) (which predicts $Q(y) \propto y S(y)$) against the
numerics in the inset of figure \ref{Fig:checkDNS}, finding a reasonably
good agreement.
Inserting (\ref{eq:ml3}) into (\ref{eq:clos2}) provides
$$
\frac{a (E_K)^{3/2}}{y} = c E_K S(y) + \beta g b^{\prime} y S(y) \frac{\alpha \Delta}{2H \ut},
$$
which can be recast, introducing the wall units and the definitions of $\Rtau$ and $Ra$, in the following form 
\begin{equation} \label{eq:clos3}
\frac{1}{\kappa y^+} = S^+\left(1+C_S \frac{Ra}{Pr^2 \Rtau^4}y^+ \right);
\end{equation}
it must be noticed that
the two parameters $\kappa$ and $C_S$ are combinations of dimensionless quantities
emerging in the derivation, which cannot be, however, derived from
first principles; fits of the numerical data indicate that good
estimates are the values $\kappa = 0.42$ and $C_S =2.5$.
Explicitating the shear term in (\ref{eq:clos3}), we get 
\begin{equation} \label{eq:clos4}
\frac{d U^+}{d y^+} = \frac{1}{\kappa y^+ (1+\kappa_C y^+)},
\end{equation}
whose integration finally yields expressions (\ref{eq:tLog-Law}) and (\ref{eq:Kc}) for the velocity profile. The 
robustness of the model is confirmed in figure \ref{Fig:tLogLawConstant} where we plot the fitted values of the parameter
 $\kappa_C$ as function of $Ra$ for fixed $\Rtau$ (main panel) and as function of $\Rtau$ for fixed $Ra$ (inset), 
together with the predictions of equation (\ref{eq:Kc}) (dashed lines).  

  Before concluding, we assimilate our results, {\it
    mutatis mutandis}, to those
  obtained in the literature for PBL. Our
  setup differs from PBL in that we simulate a channel flow with fixed
  temperature (and zero velocity) at the two walls;
  unlike the classical MO54 approach \citep{MO54,Dyer}, then, as
  written in equations (\ref{eq:ml}) and (\ref{eq:ml3}), we do
  not assume the heat flux to be constant as a sort of boundary
  condition \citep{MO54} (which is an
  approximation, as pointed out in \citet{MellorYamada2}). However,
  some analogies might be considered. With respect
  to other models \citep{Dubrulle2,MellorYamada2}, our approach enjoys
  some peculiar features: $\left. i \right)$ it relies on one single empirical input
  (the parameter $C_S$); $\left. ii \right)$ it provides an explicit analytic expression for the
  velocity profile and for its dependence on $Ra$, $Re_{\tau}$ and
  $Pr$, which makes it particularly suitable for comparison against
  data from channel flow simulations where one has direct control on
  these parameters. On the other hand, our model must not be expected
  to be fully trustable for high values of the stability
  parameter $\zeta=y/L_O$, where $L_O = u_{\tau}^3/(\kappa \beta g Q_w)$
 is the Obukhov length \citep{MO54} and $Q_w$ the heat flux at the wall. 
 For a further comparison, it is worth rewriting equation
  (\ref{eq:clos4}) in terms of the dimensionless mean
  wind gradient $\phi_U$ as 
\begin{equation} \label{eq:stability}
\phi_U(y) \equiv  \frac{\kappa y}{u_{\tau}}\frac{d U}{dy} = \left( 1
  +C_S \frac{y}{L_S} \right)^{-1},
\end{equation}
where the characteristic length $L_S$ is defined as 
\begin{equation} \label{eq:LS}
L_S = \frac{Re_{\tau}^3Pr^2}{Ra}H. 
\end{equation}
If we introduce the Nusselt number $Nu=Q_w/(\alpha \Delta/2H)$, the Obukhov length $L_O$, in turn, can be expressed in terms
of $(Ra, Re_{\tau}, Pr)$ as 
\begin{equation} \label{eq:LO}
L_O= \frac{16}{\kappa}\frac{H}{Nu}\frac{Re_{\tau}^3 Pr^2}{Ra}\approx \frac{16}{\kappa}\lambda_{\theta}\frac{Re_{\tau}^3 Pr^2}{Ra}, 
\end{equation}
where $\lambda_{\theta}$ is the boundary layer width (the last
equality is only approximated since, as mentioned before, in certain
cases there is no clear thermal shortcut \citep{GrossmannLohse}). 
A direct look at equations (\ref{eq:LS}) and (\ref{eq:LO}) suggests that
$$
L_O \propto \frac{\lambda_{\theta}}{H}L_S,
$$
but $\lambda_{\theta}$ is fixed by the heat flux from the
wall and, hence, it depends in a non-trivial way on the control
parameters. 
In the form (\ref{eq:stability}-\ref{eq:LS}), our model appears, then,
as an equivalent similarity theory for unstably stratified turbulent
channel flows.
Furthermore, the expression of the Obukhov length $L_O$ (\ref{eq:LO})
in terms of the controlling parameters allows us to check the
consistency of our data, in the logarithmic layer, with previous
theoretical studies \citep{Dubrulle2} and, by consequence, with
experimental data \citep{BusingerEtAl}. 
\begin{figure}
\begin{center}
\includegraphics[width=0.6\textwidth]{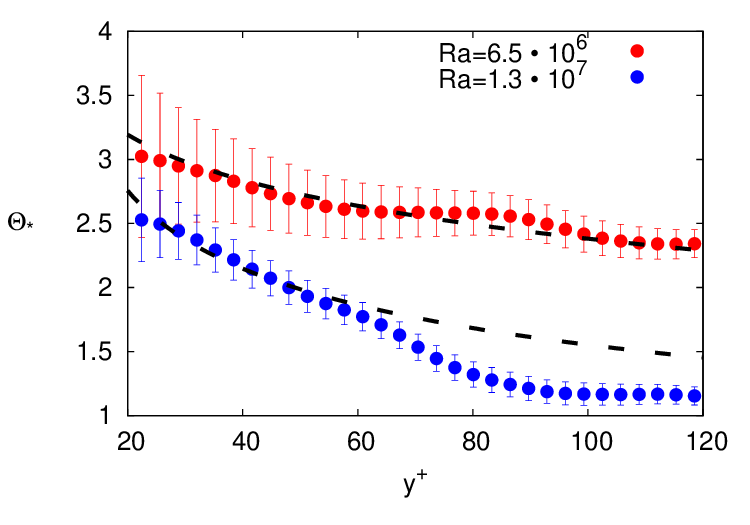}
\caption{Temperature fluctuations $\Theta_* \equiv \overline{(T -
  \overline{T})^2}^{1/2}/(Q_w/u_{\tau})$ from our LB simulations (symbols) and
RDT prediction (\ref{eq:tvarRDT}), with the expression (\ref{eq:LO})
for $L_O$,  (dashed lines) as function of the
wall-normal direction coordinate in wall units (the range shown is
restricted to the logarithmic layer). The data have been fitted with
the theoretical curves using for the parameters $(\alpha_1,\alpha_2)$
the values: ($0.26$,$0.6$), for $Ra=6.5 \times 10^6$, and
($0.22$,$0.9$), for $Ra=1.3 \times 10^7$.}
\label{Fig:tempvar}
\end{center}
\end{figure}
 In figure \ref{Fig:tempvar} we
compare the temperature fluctuations $\Theta_* \equiv \overline{(T -
  \overline{T})^2}^{1/2}/(Q_w/u_{\tau})$ from our simulations with the
RDT prediction \citep{Dubrulle2}:
\begin{equation} \label{eq:tvarRDT}
\Theta_*^{(RDT)}(\zeta) = \frac{1}{(\alpha_1 + \alpha_2 \zeta)}.
\end{equation}
The numerical data and the theoretical result (dashed line) appear to
be in fair agreement, within error bars; in particular, the dependence
on $(Ra, Re_{\tau})$ as in (\ref{eq:LO}) is fulfilled.
\section{Conclusions}
We have studied, by means of direct numerical simulations based on a thermal
lattice Boltzmann algorithm, the dynamics of a turbulent channel flow
under a gravity field orthogonal to the streamwise direction coupled
to an imposed temperature difference between the top (cold) wall
and the bottom (hot) wall. The resulting unstably stratified
configuration flattens the velocity profile and decreases the
centreline value when the buoyancy strength is increased. This
effective drag shows up in an enhancement of the friction coefficient. 
The action of buoyancy on the boundary layer structure has also been
probed looking at other relevant statistical quantities in wall
bounded turbulent system, such as Reynolds stress; we have found that, as
the Rayleigh number is increased, the squared wall normal velocity
grows in the bulk becoming comparable or even larger than the
streamwise component (which, in turn, is depleted close to the
wall). To provide a quantitative interpretation of the numerical
findings, we have proposed a phenomenological model resulting in a
modified logarithmic law of the boundary layer; such model could
successfully capture the various velocity profiles at changing the
shear Reynolds and Rayleigh number, with just one adjustable
parameter. \\
AS, HE and AG acknowledge
financial support from the Icelandic Research Fund.
AS acknowledges funding from the European Research Council under the EU
Seventh Framework Programme (FP7/2007-2013) / ERC Grant Agreement no[279004].
This work was partially supported by the Foundation for Fundamental
Research on Matter (FOM), a part of the Netherlands Organisation for
Scientific Research (NWO), and by the COST action MP1305.

\bibliographystyle{jfm}

\end{document}